\PassOptionsToPackage{unicode}{hyperref}
\PassOptionsToPackage{hyphens}{url}
\documentclass[
]{article}
\usepackage{xcolor}
\usepackage{amsmath,amssymb}
\usepackage{iftex}
\ifPDFTeX
  \usepackage[T1]{fontenc}
  \usepackage[utf8]{inputenc}
  \usepackage{textcomp} 
\else 
  \usepackage{unicode-math} 
  \defaultfontfeatures{Scale=MatchLowercase}
  \defaultfontfeatures[\rmfamily]{Ligatures=TeX,Scale=1}
\fi
\usepackage{lmodern}
\ifPDFTeX\else
\fi
\IfFileExists{upquote.sty}{\usepackage{upquote}}{}
\IfFileExists{microtype.sty}{
  \usepackage[]{microtype}
  \UseMicrotypeSet[protrusion]{basicmath} 
}{}
\makeatletter
\@ifundefined{KOMAClassName}{
  \IfFileExists{parskip.sty}{%
    \usepackage{parskip}
  }{
    \setlength{\parindent}{0pt}
    \setlength{\parskip}{6pt plus 2pt minus 1pt}}
}{
  \KOMAoptions{parskip=half}}
\makeatother
\usepackage{color}
\usepackage{fancyvrb}

\DefineVerbatimEnvironment{Highlighting}{Verbatim}{commandchars=\\\{\}}
\newenvironment{Shaded}{}{}

\newcommand{\AttributeTok}[1]{\textcolor[rgb]{0.49,0.56,0.16}{#1}}

\newcommand{\DataTypeTok}[1]{\textcolor[rgb]{0.56,0.13,0.00}{#1}}
\newcommand{\DecValTok}[1]{\textcolor[rgb]{0.25,0.63,0.44}{#1}}

\newcommand{\ExtensionTok}[1]{#1}

\newcommand{\FunctionTok}[1]{\textcolor[rgb]{0.02,0.16,0.49}{#1}}

\newcommand{\KeywordTok}[1]{\textcolor[rgb]{0.00,0.44,0.13}{\textbf{#1}}}
\newcommand{\NormalTok}[1]{#1}

\newcommand{\OtherTok}[1]{\textcolor[rgb]{0.00,0.44,0.13}{#1}}

\newcommand{\StringTok}[1]{\textcolor[rgb]{0.25,0.44,0.63}{#1}}

\usepackage{longtable,booktabs,array}
\usepackage{calc} 
\usepackage{etoolbox}
\makeatletter
\patchcmd\longtable{\par}{\if@noskipsec\mbox{}\fi\par}{}{}
\makeatother
\IfFileExists{footnotehyper.sty}{\usepackage{footnotehyper}}{\usepackage{footnote}}
\makesavenoteenv{longtable}
\usepackage{graphicx}
\makeatletter
\newsavebox\pandoc@box
\newcommand*\pandocbounded[1]{
  \sbox\pandoc@box{#1}%
  \Gscale@div\@tempa{\textheight}{\dimexpr\ht\pandoc@box+\dp\pandoc@box\relax}%
  \Gscale@div\@tempb{\linewidth}{\wd\pandoc@box}%
  \ifdim\@tempb\p@<\@tempa\p@\let\@tempa\@tempb\fi
  \ifdim\@tempa\p@<\p@\scalebox{\@tempa}{\usebox\pandoc@box}%
  \else\usebox{\pandoc@box}%
  \fi%
}
\def\fps@figure{htbp}
\makeatother
\providecommand{\tightlist}{%
  \setlength{\itemsep}{0pt}\setlength{\parskip}{0pt}}
\usepackage[]{natbib}
\usepackage{bookmark}
\IfFileExists{xurl.sty}{\usepackage{xurl}}{} 
\hypersetup{
  pdftitle={NBI-Slurm: Simplified submission of Slurm jobs with energy saving mode},
  pdfauthor={Andrea Telatin Quadram Institute Bioscience, Norwich, NR4 7UQ, UK Center for Microbial Interactions, Norwich Research Park, Norwich, NR4 7UG, UK ORCID: 0000-0001-7619-281X},
  hidelinks,
  pdfcreator={LaTeX via pandoc}}

\title{NBI-Slurm: Simplified submission of Slurm jobs with energy saving
mode}
\author{Andrea Telatin\\
Quadram Institute Bioscience, Norwich, NR4 7UQ, UK\\
Center for Microbial Interactions, Norwich Research Park, Norwich, NR4 7UG, UK\\
ORCID: 0000-0001-7619-281X}
\date{18 March 2026}

\begin{document}
\maketitle

\section{Summary}\label{summary}

NBI-Slurm is a Perl package that provides a simplified, user-friendly
interface for submitting and managing jobs on SLURM
\citep{jette2002slurm} high-performance computing (HPC) clusters. It
offers both a library of Perl modules for programmatic job management
and a suite of command-line tools designed to reduce the cognitive
overhead of SLURM's native interface. Distinctive features of NBI-Slurm
are (a) TUI applications to view and cancel jobs, (b) the possibility to
generate tool specific wrappers for (bioinformatic) tools and (c) an
energy-aware scheduling mode --- ``eco mode'' --- that automatically
defers flexible jobs to off-peak periods, helping research institutions
reduce their computational carbon footprint without requiring users to
manually plan submission times.

\section{Statement of Need}\label{statement-of-need}

HPC clusters are indispensable in modern research, particularly in the
life sciences where large-scale sequence analyses, genome assemblies,
and statistical models demand resources beyond a desktop workstation.
SLURM has become the dominant workload manager in this space
\citep{slurm_adoption}, yet its interface presents a steep learning
curve. Users must learn a verbose \texttt{sbatch} scripting syntax,
understand resource unit conventions (memory in megabytes, time in
\texttt{D-HH:MM:SS} format), manage job dependencies manually, and
repeat boilerplate directives across every submission script.

Workflow managers such as Snakemake \citep{molder2021snakemake} and
Nextflow \citep{di2017nextflow} address this at the pipeline level by
abstracting SLURM as an execution backend, but they require users to
rewrite their analysis logic inside a domain-specific language. Many
researchers have existing shell scripts or one-off analyses that do not
warrant a full pipeline refactor. NBI-Slurm occupies a complementary
niche: it wraps SLURM's interface without imposing a workflow model,
making it straightforward to submit individual commands or small batches
while retaining access to all SLURM features through pass-through
options.

\begin{figure}
\centering
\pandocbounded{\includegraphics[keepaspectratio,alt={Interactive TUI of viewjobs, showing job navigation, multi-column display, and bulk-cancel workflow. The image is AI generated from a real screenshot (Google NanoBanana) }]{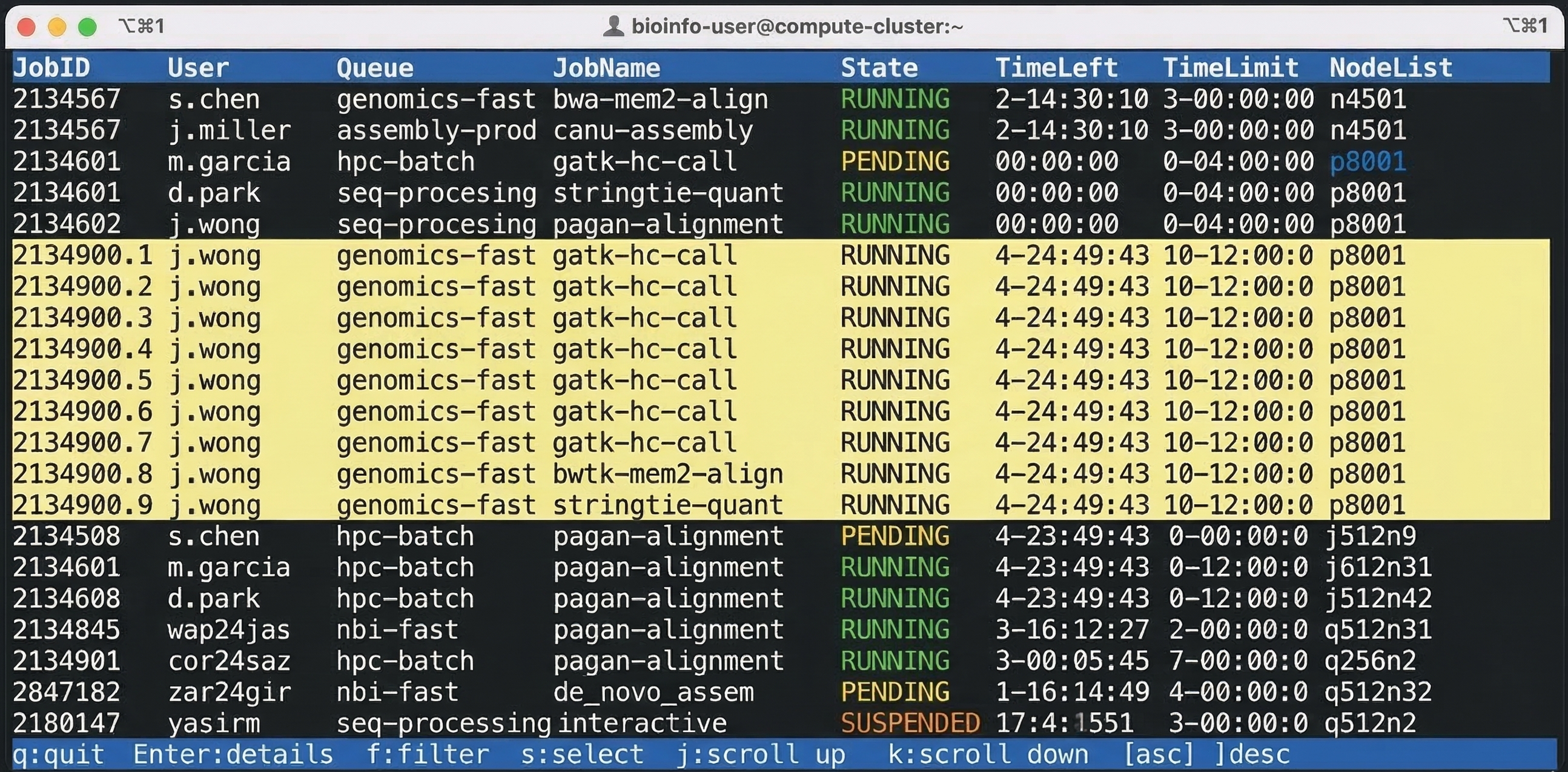}}
\caption{Interactive TUI of \protect\texttt{viewjobs}, showing job
navigation, multi-column display, and bulk-cancel workflow. The image is
AI generated from a real screenshot (Google NanoBanana)
\label{fig:viewjobs}}
\end{figure}

The \texttt{lsjobs} utility prints a colour-coded, human-readable table
of queued jobs as a static snapshot, offering a more ergonomic
alternative to the raw output of \texttt{squeue}. Its companion tool
\texttt{viewjobs} provides a fully interactive terminal user interface
(TUI) that allows users to browse the live job queue without leaving the
terminal (\autoref{fig:viewjobs}). Users can scroll through jobs with
arrow or Vim keys, sort columns, inspect per-job details, toggle column
visibility, and adjust column widths interactively. Individual jobs can
be selected with \texttt{Space} and multiple selected jobs can be
cancelled in bulk with a single keypress, removing the need to
copy-paste job IDs into \texttt{scancel}.

Energy consumption in research computing is a growing concern
\citep{lannelongue2021green}. Most researchers have no practical
mechanism to shift flexible jobs to periods when grid electricity is
cheaper or cleaner. NBI-Slurm addresses this directly with a
configurable scheduling module that calculates the next available
low-energy window and injects a \texttt{-\/-begin} directive into the
submission, requiring no change to the underlying command.

\section{NBI::Slurm package}\label{nbislurm-package}

\subsection{Availability and
Installation}\label{availability-and-installation}

NBI-Slurm is distributed under the MIT licence and is available from
CPAN as \texttt{NBI::Slurm}. Installation requires Perl 5.16 or later
and can be performed with:

\begin{Shaded}
\begin{Highlighting}[]
\ExtensionTok{cpanm}\NormalTok{ NBI::Slurm}
\end{Highlighting}
\end{Shaded}

The source code is hosted at
\url{https://github.com/quadram-institute-bioscience/NBI-Slurm} under
continuous integration. Development has been active since June 2023, and
the module is published to the MetaCPAN repository at
\url{https://metacpan.org/dist/NBI-Slurm}.

\subsection{Code Structure and
Dependencies}\label{code-structure-and-dependencies}

The package is organised into two layers.

\textbf{Perl module library (\texttt{lib/NBI/}):} Five classes model the
key abstractions:

\begin{itemize}
\tightlist
\item
  \texttt{NBI::Opts} --- encapsulates SLURM resource directives (queue,
  threads, memory, wall-time, email, job arrays, start time). It accepts
  human-friendly inputs such as \texttt{"8GB"} or \texttt{"2h30m"} and
  converts them to SLURM's expected formats.
\item
  \texttt{NBI::Job} --- represents a job to be submitted, holding a
  command (or list of commands) and an \texttt{NBI::Opts} object. The
  \texttt{script()} method generates a complete \texttt{sbatch} script;
  \texttt{run()} submits it and returns the job identifier.
\item
  \texttt{NBI::Queue} --- queries the live SLURM queue via
  \texttt{squeue} and returns a list of \texttt{NBI::QueuedJob} objects,
  optionally filtered by user, status, name, or queue.
\item
  \texttt{NBI::QueuedJob} --- a lightweight data object representing one
  queued job, used by the queue-management tools.
\item
  \texttt{NBI::EcoScheduler} --- implements the energy-aware scheduling
  logic. Given a job's expected duration and a set of configurable
  windows, it finds the next period satisfying a three-tier preference:
  (1) job completes within an eco window and avoids peak hours; (2) job
  starts in an eco window and avoids peak hours but may overrun; (3) job
  starts in an eco window and partially overlaps peak hours. Default
  windows target weekday nights (00:00--06:00) and weekend off-peak
  periods (00:00--07:00, 11:00--16:00), avoiding evening peaks
  (17:00--20:00), all of which are fully configurable with a settings
  file (by default \texttt{\textasciitilde{}/.nbislurm.config}).
\end{itemize}

\textbf{Command-line tools (\texttt{bin/}):}

{\def\LTcaptype{none} 
\begin{longtable}[]{@{}
  >{\raggedright\arraybackslash}p{(\linewidth - 2\tabcolsep) * \real{0.4000}}
  >{\raggedright\arraybackslash}p{(\linewidth - 2\tabcolsep) * \real{0.6000}}@{}}
\toprule\noalign{}
\begin{minipage}[b]{\linewidth}\raggedright
Tool
\end{minipage} & \begin{minipage}[b]{\linewidth}\raggedright
Purpose
\end{minipage} \\
\midrule\noalign{}
\endhead
\bottomrule\noalign{}
\endlastfoot
\texttt{runjob} & Submit a command as a SLURM job with resource flags \\
\texttt{lsjobs} & List, filter, and cancel user jobs with coloured
tabular output \\
\texttt{viewjobs} & Interactive terminal UI for job management \\
\texttt{waitjobs} & Block until jobs matching a pattern complete \\
\texttt{whojobs} & Show cluster utilisation grouped by user \\
\texttt{session} & Launch an interactive SLURM session \\
\end{longtable}
}

Runtime dependencies are deliberately minimal: \texttt{Capture::Tiny}
(\textgreater=0.40), \texttt{JSON::PP}, \texttt{Text::ASCIITable}
(\textgreater=0.22), \texttt{Term::ANSIColor}, \texttt{Storable}, and
\texttt{POSIX}---all either part of the Perl core or widely available on
CPAN.

\subsection{Documentation}\label{documentation}

Each module is documented with embedded POD (Plain Old Documentation),
rendered on CPAN at \url{https://metacpan.org/dist/NBI-Slurm}. Each
command-line tool provides a \texttt{-\/-help} flag and a manual page
generated from its POD. A user guide with annotated examples is
maintained in the repository's \texttt{README.md}. The test suite
(\texttt{t/}) covers unit behaviour of every module and integration
behaviour of the command-line tools; author-facing tests (\texttt{xt/})
verify POD completeness and coverage. All tests will be able to check
functions even without Slurm. To check the ability to interact with
Slurm, there are optional tests that can be executed with
\texttt{prove\ -lv\ xt/hpc-*.t}.

\subsection{Wrappers}\label{wrappers}

NBI-Slurm includes a declarative wrapper framework built around three
classes: \texttt{NBI::Launcher}, \texttt{NBI::Manifest}, and
\texttt{NBI::Pipeline}. A wrapper is a small Perl module that subclasses
\texttt{NBI::Launcher} and describes a bioinformatics tool --- its
inputs, parameters, outputs, activation method (HPC module, conda
environment, or Singularity image), and SLURM resource defaults --- in a
single constructor call. The only method that subclasses typically need
to override is \texttt{make\_command()}, which returns the tool
invocation string; the base class handles input validation,
scratch-directory setup, shell script generation, and job submission.
\texttt{NBI::Manifest} serialises all resolved inputs, parameters,
outputs, and SLURM resources to a JSON provenance file written alongside
the results at submission time, then patched in-place by the job script
itself upon completion or failure --- with no dependency on external
tools such as \texttt{jq}. Multi-step analyses can be expressed as
\texttt{NBI::Pipeline} objects that wire \texttt{afterok} SLURM
dependencies between \texttt{NBI::Job} instances automatically. The
bundled \texttt{NBI::Launcher::Kraken2} module illustrates the pattern:
it declares paired- or single-end FASTQ inputs, a database directory
that defaults to the \texttt{KRAKEN2\_DB} environment variable, and a
\texttt{threads} parameter that is automatically synchronised from the
\texttt{-\/-cpus} SLURM flag; its \texttt{build()} override measures the
database folder size at submission time and inflates the memory request
accordingly (40\% headroom plus a 100 GB fixed overhead), ensuring the
job is unlikely to be killed by the out-of-memory handler without
requiring the user to perform any calculation. Third-party wrappers can
be placed in \texttt{\textasciitilde{}/.nbi/launchers/} and are
discovered automatically by the \texttt{nbilaunch} command-line tool.

\subsection{Example commands}\label{example-commands}

\textbf{Submitting a parallel job.} A researcher wishing to run a genome
assembler with 18 cores, 64 GB RAM, and a 12-hour wall-time can write:

\begin{Shaded}
\begin{Highlighting}[]
\ExtensionTok{runjob} \AttributeTok{{-}n} \StringTok{"assembly"} \AttributeTok{{-}c}\NormalTok{ 18 }\AttributeTok{{-}m}\NormalTok{ 64 }\AttributeTok{{-}t}\NormalTok{ 12 }\AttributeTok{{-}w}\NormalTok{ ./logs/ }\DataTypeTok{\textbackslash{}}
  \StringTok{"flye {-}{-}nano{-}raw reads.fastq {-}{-}out{-}dir asm"}
\end{Highlighting}
\end{Shaded}

\textbf{Processing a file list as a job array.} To align 200 FASTQ
files, one job per file:

\begin{Shaded}
\begin{Highlighting}[]
\ExtensionTok{runjob} \AttributeTok{{-}n} \StringTok{"align"} \AttributeTok{{-}c}\NormalTok{ 8 }\AttributeTok{{-}m}\NormalTok{ 16 }\AttributeTok{{-}{-}files}\NormalTok{ samples.txt }\DataTypeTok{\textbackslash{}}
  \StringTok{"bwa mem ref.fa \#FILE\# \textgreater{} \#FILE\#.bam"}
\end{Highlighting}
\end{Shaded}

\textbf{Energy-aware deferral.} A long-running but flexible annotation
job can be scheduled for the next eco window automatically. Note that by
default the \emph{eco mode} is enabled, and can be overridden with
\texttt{-\/-no-eco} or setting the economy\_mode=0 in the configuration
file.

\begin{Shaded}
\begin{Highlighting}[]
\ExtensionTok{runjob} \AttributeTok{{-}{-}eco} \AttributeTok{{-}n} \StringTok{"annotate"} \AttributeTok{{-}t}\NormalTok{ 6 }\StringTok{"prokka genome.fa"}
\end{Highlighting}
\end{Shaded}

NBI-Slurm calculates the next suitable window (e.g., the following
night) and adds \texttt{-\/-begin=2026-03-19T00:00:00} to the submission
without any further user action.

\textbf{Programmatic job chaining.} In a Perl analysis script:

\begin{Shaded}
\begin{Highlighting}[]
\FunctionTok{use} \FunctionTok{NBI::Job}\NormalTok{;}
\FunctionTok{use} \FunctionTok{NBI::Opts}\NormalTok{;}

\KeywordTok{my} \DataTypeTok{$opts}\NormalTok{ = }\FunctionTok{NBI::Opts}\NormalTok{{-}\textgreater{}new(}
\NormalTok{    {-}queue =\textgreater{} }\OtherTok{"}\StringTok{long}\OtherTok{"}\NormalTok{, }
\NormalTok{    {-}threads =\textgreater{} }\DecValTok{16}\NormalTok{, }
\NormalTok{    {-}memory =\textgreater{} }\DecValTok{32}\NormalTok{, }
\NormalTok{    {-}}\FunctionTok{time}\NormalTok{ =\textgreater{} }\OtherTok{"}\StringTok{4h}\OtherTok{"}\NormalTok{);}
\KeywordTok{my} \DataTypeTok{$job}\NormalTok{  = }\FunctionTok{NBI::Job}\NormalTok{{-}\textgreater{}new(}
\NormalTok{    {-}name =\textgreater{} }\OtherTok{"}\StringTok{step1}\OtherTok{"}\NormalTok{, }
\NormalTok{    {-}command =\textgreater{} }\OtherTok{"}\StringTok{python align.py}\OtherTok{"}\NormalTok{, }
\NormalTok{    {-}opts =\textgreater{} }\DataTypeTok{$opts}\NormalTok{);}
\KeywordTok{my} \DataTypeTok{$id}\NormalTok{   = }\DataTypeTok{$job}\NormalTok{{-}\textgreater{}}\DataTypeTok{run}\NormalTok{();}

\KeywordTok{my} \DataTypeTok{$opts2}\NormalTok{ = }\FunctionTok{NBI::Opts}\NormalTok{{-}\textgreater{}new(}
\NormalTok{    {-}queue =\textgreater{} }\OtherTok{"}\StringTok{short}\OtherTok{"}\NormalTok{, }
\NormalTok{    {-}threads =\textgreater{} }\DecValTok{4}\NormalTok{, }
\NormalTok{    {-}memory =\textgreater{} }\DecValTok{8}\NormalTok{, }
\NormalTok{    {-}}\FunctionTok{time}\NormalTok{ =\textgreater{} }\OtherTok{"}\StringTok{1h}\OtherTok{"}\NormalTok{);}
\KeywordTok{my} \DataTypeTok{$job2}\NormalTok{  = }\FunctionTok{NBI::Job}\NormalTok{{-}\textgreater{}new(}
\NormalTok{    {-}name =\textgreater{} }\OtherTok{"}\StringTok{step2}\OtherTok{"}\NormalTok{, }
\NormalTok{    {-}command =\textgreater{} }\OtherTok{"}\StringTok{python report.py {-}{-}input results/}\OtherTok{"}\NormalTok{, }
\NormalTok{    {-}opts =\textgreater{} }\DataTypeTok{$opts2}\NormalTok{);}
\DataTypeTok{$job2}\NormalTok{{-}\textgreater{}}\DataTypeTok{opts}\NormalTok{{-}\textgreater{}}\DataTypeTok{dependencies}\NormalTok{([}\DataTypeTok{$id}\NormalTok{]);}
\DataTypeTok{$job2}\NormalTok{{-}\textgreater{}}\DataTypeTok{run}\NormalTok{();}
\end{Highlighting}
\end{Shaded}

\section{Acknowledgements}\label{acknowledgements}

The author gratefully acknowledges the support of the Biotechnology and
Biological Sciences Research Council (BBSRC); this research was funded
by the BBSRC Institute Strategic Programme Food Microbiome and Health
BB/X011054/1 and its constituent project(s) BBS/E/QU/230001B; the BBSRC
Institute Strategic Programme Microbes and Food Safety BB/X011011/1 and
its constituent project(s) BBS/E/QU/230002C; the BBSRC Core Capability
Grant BB/CCG2260/1. This research was also supported by the
infrastructure provided by the CLIMB-BIG-DATA grant MR/T030062/1. The
author thanks colleagues at the Quadram Institute Bioscience for
feedback and field-testing during development, and the GreenDISC working
group and NBI Research Computing for support and discussions.

\section{AI Usage Disclosure}\label{ai-usage-disclosure}

Claude Code (Anthropic) was used during development of NBI-Slurm from
version 0.10.0 onwards, assisting with code generation, refactoring,
test scaffolding, and documentation drafting. It was also used to assist
with drafting and editing this paper. All AI-assisted outputs were
reviewed, edited, and validated by the author, who made all core design
decisions and retains full responsibility for the accuracy, originality,
and correctness of the submitted materials.

\renewcommand\refname{References}
\bibliography{paper.bib}

\begin{thebibliography}{5}
\providecommand{\natexlab}[1]{#1}
\providecommand{\url}[1]{\texttt{#1}}
\expandafter\ifx\csname urlstyle\endcsname\relax
  \providecommand{\doi}[1]{doi: #1}\else
  \providecommand{\doi}{doi: \begingroup \urlstyle{rm}\Url}\fi

\bibitem[Di~Tommaso et~al.(2017)Di~Tommaso, Chatzou, Floden, Barja, Palumbo,
  and Notredame]{di2017nextflow}
Paolo Di~Tommaso, Maria Chatzou, Evan~W Floden, Pablo~Prieto Barja, Emilio
  Palumbo, and Cedric Notredame.
\newblock Nextflow enables reproducible computational workflows.
\newblock \emph{Nature Biotechnology}, 35\penalty0 (4):\penalty0 316--319,
  2017.
\newblock \doi{10.1038/nbt.3820}.

\bibitem[Lannelongue et~al.(2021)Lannelongue, Grealey, and
  Inouye]{lannelongue2021green}
Loïc Lannelongue, Jason Grealey, and Michael Inouye.
\newblock Green algorithms: Quantifying the carbon footprint of computation.
\newblock \emph{Advanced Science}, 8\penalty0 (12), May 2021.
\newblock ISSN 2198-3844.
\newblock \doi{10.1002/advs.202100707}.
\newblock URL \url{http://dx.doi.org/10.1002/advs.202100707}.

\bibitem[M\"{o}lder et~al.(2021)M\"{o}lder, Jablonski, Letcher, Hall,
  Tomkins-Tinch, Sochat, Forster, Lee, Twardziok, Kanitz, Wilm, Holtgrewe,
  Rahmann, Nahnsen, and K\"{o}ster]{molder2021snakemake}
Felix M\"{o}lder, Kim~Philipp Jablonski, Brice Letcher, Michael~B. Hall,
  Christopher~H. Tomkins-Tinch, Vanessa Sochat, Jan Forster, Soohyun Lee,
  Sven~O. Twardziok, Alexander Kanitz, Andreas Wilm, Manuel Holtgrewe, Sven
  Rahmann, Sven Nahnsen, and Johannes K\"{o}ster.
\newblock Sustainable data analysis with snakemake.
\newblock \emph{F1000Research}, 10:\penalty0 33, April 2021.
\newblock ISSN 2046-1402.
\newblock \doi{10.12688/f1000research.29032.2}.
\newblock URL \url{http://dx.doi.org/10.12688/f1000research.29032.2}.

\bibitem[Wang et~al.(2020)Wang, Chen, and Xiao]{slurm_adoption}
Bo~Wang, Zhiguang Chen, and Nong Xiao.
\newblock A survey of system scheduling for hpc and big data.
\newblock In \emph{Proceedings of the 2020 4th International Conference on High
  Performance Compilation, Computing and Communications}, HP3C 2020, page
  178–183. ACM, June 2020.
\newblock \doi{10.1145/3407947.3407977}.
\newblock URL \url{http://dx.doi.org/10.1145/3407947.3407977}.

\bibitem[Yoo et~al.(2003)Yoo, Jette, and Grondona]{jette2002slurm}
Andy~B. Yoo, Morris~A. Jette, and Mark Grondona.
\newblock Slurm: Simple linux utility for resource management.
\newblock In \emph{Job Scheduling Strategies for Parallel Processing}, page
  44–60. Springer Berlin Heidelberg, 2003.
\newblock ISBN 9783540397274.
\newblock \doi{10.1007/10968987_3}.
\newblock URL \url{http://dx.doi.org/10.1007/10968987_3}.

\end{thebibliography}

\end{document}